# Overview of the Tevatron Collider Complex: Goals, Operations and Performance [1]

**Stephen Holmes**[a], **Ronald S. Moore**[a*]**, and Vladimir Shiltsev**[a]

[a] *Fermi National Accelerator Laboratory*
*PO Box 500, Batavia, IL, 60510, USA*
*E-mail*: `ronmoore@fnal.gov`

ABSTRACT: For more than two decades the Tevatron proton-antiproton collider was the centerpiece of the world's high energy physics program. The collider was arguably one of the most complex research instruments ever to reach the operation stage and is widely recognized for numerous physics discoveries and for many technological breakthroughs. In this article we outline the historical background that led to the construction of the Tevatron Collider, the strategy applied to evolution of performance goals over the Tevatron's operational history, and briefly describe operations of each accelerator in the chain and achieved performance.

---

[1] Work supported by Fermi Research Alliance, LLC under Contract No. De-AC02-07CH11359 with the United States Department of Energy.
* Corresponding author

# Contents

## 1. Tevatron Collider Runs I and II: Goals and Strategy

### 1.1 Historic background

The Tevatron was conceived to double the energy of the Fermilab complex from 500 GeV to 1000 GeV. The original name, the "Energy Saver/Doubler", reflected this mission and the accrued benefit of reduced power utilization through the use of superconducting magnets [1]. The introduction of superconducting magnets in a large scale application allowed the (now named) Tevatron to be constructed with the same circumference, and to be installed in the same tunnel, as the original Main Ring which would serve as its injector (at 150 GeV). Superconducting magnet development was initiated in the early 1970's and ultimately produced successful magnets, leading to commissioning of the Tevatron in July 1983.

In 1976 Carlo Rubbia, David Cline, Peter McIntyre and colleagues proposed the implementation of a proton-antiproton collider at Fermilab [2] or at CERN, based on the conversion of an existing accelerator into a storage ring and construction of a new facility for the accumulation and cooling of antiprotons. The motivation was to discover the intermediate vector bosons, and the proposal argued that the requisite luminosity ($\sim 10^{29}$ cm$^{-2}$sec$^{-1}$) could be achieved with a facility that would produce and cool approximately $10^{11}$ antiprotons per day. The first antiproton accumulation facility was constructed at CERN and supported collisions at a center-of-mass energy $\sqrt{s}$ = 630 GeV in the modified SPS synchrotron. Operations commenced in 1981 and led to the discovery of the W and Z particles in 1983. Meanwhile, in 1978 Fermilab decided that proton-antiproton collisions would be supported in the Tevatron, at roughly $\sqrt{s}$ = 2000 GeV and that a facility would be constructed to supply the required antiprotons. The Tevatron I Project was established to design and manage the construction of the Fermilab Antiproton Source and the conversions required for colliding beams in the Tevatron. Design goals were established as a luminosity of $1\times10^{30}$ cm$^{-2}$sec$^{-1}$ at a center-of-mass energy $\sqrt{s} = 1800$ GeV.

Funding for Tevatron I was initiated in 1981 and the Tevatron was completed, as a fixed target accelerator, in the summer of 1983. The Antiproton Source was completed in 1985 and first collisions were observed in the Tevatron using operational elements of the CDF detector (then under construction) in October 1985. Initial operations of the collider for data taking took place over the period February – May of 1987. A more extensive run took place over June 1988 – June 1989, representing the first sustained operations at the design luminosity. Over that period a total of 5 pb$^{-1}$ were delivered to CDF at 1800 GeV (center-of-mass) and the first western hemisphere W's and Z's were observed. The initial operational goal of $1\times10^{30}$ cm$^{-2}$sec$^{-1}$ luminosity was exceeded during this run. Table I summarizes the goals established during the Tevatron I design phase and the actual performance achieved in the 1988-89 run. (A short run at $\sqrt{s} = 1020$ GeV also occurred in 1989.)

Table I: Design and achieved performance parameters for the 1988-89 Collider Run (typical values at the beginning of a store).

|  | Tevatron I Design | 1988-89 Actual |  |
|---|---|---|---|
| Energy (center-of-mass) | 1800 | 1800 | GeV |
| Protons/bunch, $N_p$ | $6\times10^{10}$ | $7.0\times10^{10}$ |  |
| Antiprotons/bunch, $N_a$ | $6\times10^{10}$ | $2.9\times10^{10}$ |  |
| Number of bunches, $n_b$ | 3 | 6 |  |
| Total Antiprotons, $n_b N_a$ | $18\times10^{10}$ | $17\times10^{10}$ |  |
| Proton emittance (rms, normalized), $\varepsilon_{pn}$ | 3.3 | 4.2 | $\pi$ mm-mrad |
| Antiproton emittance (rms, normalized), $\varepsilon_{an}$ | 3.3 | 3 | $\pi$ mm-mrad |
| IP beta-function, $\beta^*$ | 100 | 55 | cm |
| Luminosity | $1\times10^{30}$ | $1.6\times10^{30}$ | cm$^{-2}$sec$^{-1}$ |

**1.2 Performance limitations**

The luminosity achievable in a proton-antiproton collider can be written as:

$$L = \frac{f_0 n_b N_p N_a}{2\pi} \frac{1}{\sigma_p^2 + \sigma_a^2} H\left(\frac{\sigma_z}{\beta^*}\right) = \frac{\gamma f_0 (N_p / \varepsilon_{pn})(n_b N_a)}{2\pi\beta^*} \frac{1}{1 + \varepsilon_{an}/\varepsilon_{pn}} H\left(\frac{\sigma_z}{\beta^*}\right) \quad (1)$$

The luminosity formula is written in this form to highlight the limitations inherent in operations of the Tevatron:

- $N_p/\varepsilon_{pn}$ is the number of protons per bunch divided by the beam emittance. This quantity is directly proportional to the beam-beam tune shift $\xi_a = r_p N_p/4\pi\varepsilon_{pn}$ experienced by the antiprotons for each head-on encounter with the protons. With six bunch operations there are a total of 12 such encounters per revolution. During the 1988-89 Run it was observed that the total available tune shift that could be tolerated was determined by the tune space available between resonances below about $10^{th}$ order. The Tevatron was operated between the 2/5 and 3/7 resonances, allowing a total tune shift (summed over the 12 encounters) of about 0.028, similar to what had been obtained in the Sp$\bar{p}$S. It was observed that the Tevatron could operate in this tune region with no deleterious impacts from the $12^{th}$ order resonance lying in between the $5^{th}$ and $7^{th}$ resonances.

- $n_b N_a$ is the total number of antiprotons in the collider. Two limitations existed: the antiproton accumulation rate, and the ability to cool and store a suitably large number of antiprotons for delivery to the Tevatron. The antiproton accumulation rate is dictated by the rate at which protons can be delivered to the antiproton production target, and by the aperture and stochastic cooling capabilities of the Antiproton Source (see Section 2.4). The accumulation rate was roughly $2 \times 10^{10}$ antiprotons/hour during the 1988-89 Run based on $2 \times 10^{12}$ protons delivered to the antiproton production target every 2.6 seconds. The total number of antiprotons that could be delivered to the Tevatron was determined by an interplay between the available Main Ring aperture and the correlation between emittance and antiproton intensity imposed by the stochastic cooling systems in the Antiproton Source. It was determined that for the emittance that could fit through the Main Ring aperture the maximum antiproton intensity in the Antiproton Source was about $6 \times 10^{11}$. At this level the Main Ring transmission efficiency for antiprotons was about 60%.

- $\gamma = E_p/m_p c^2$ is a measure of the beam energy. For all other parameters fixed, the luminosity is proportional to the beam energy, so any increase in beam energy will increase the luminosity. In addition beam energy increases have the added benefit of increasing the cross sections for the production of high mass states.

- The luminosity is inversely proportional to the beta-function at the interaction point, so any decrease in the $\beta^*$ improves the luminosity. The impact of this is ultimately limited by the form factor, also called hourglass factor, $H(\sigma_z/\beta^*)$ when the ratio of the rms bunch length to the beta function $\sigma_z/\beta^*$ becomes larger than 1.

**1.3 Strategy**

Based on the above considerations a long term strategy was developed in the late 1980's for Tevatron upgrades to a luminosity of $5 \times 10^{31}$ cm$^{-2}$sec$^{-1}$, a factor of 50 beyond the original goal [3]. The primary elements of the strategy were:

- <u>Electrostatic separators</u>: Twenty-two, 3 m long, electrostatic separators operating at up to ±300 kV across a 5 cm gap were installed into the Tevatron. The separator installation allowed operations with protons and antiprotons traveling on separated helical orbits in the Tevatron. This installation was aimed at mitigating the beam-beam limitations (the head-on collisions would now happen only in the two interaction points located inside the CDF and D0 particle physics detectors) and allowed an increase in the number of bunches (thus keeping the interactions/crossing seen in the detectors under control) as the luminosity increased. As a result of operations with separated orbits the beam-beam effect ceased to be a limitation even as the proton intensity and number of bunches were increased. During Run II, 4 additional separators were installed to improve separation at the nearest parasitic crossings.

- Low beta systems: The 1988-89 Run did not have a matched insertion for the interaction region at B0 (where CDF was situated). Two sets of high performance quadrupoles were developed and installed at B0 and D0 (which came online for Run I in 1992). These systems ultimately allowed operations with $\beta^*$ less than 30 cm at the end of the collider Run II.
- Cryogenic cold compressors: Cryogenic cold compressors were introduced into the Tevatron to lower the operating temperature by about 0.5 K, thereby allowing the beam energy to be increased to 1000 GeV, in theory. In operational practice 980 GeV was achieved.
- 400 MeV Linac Upgrade: The 200 MeV linac was upgraded to 400 MeV to reduce space-charge effects at injection energy in Booster which provided higher beam brightness at 8 GeV. The total intensity delivered from the Booster increased from roughly $3\times10^{12}$ per pulse to about $5\times10^{12}$. This resulted in more protons being transmitted to the antiproton production target and, ultimately, more protons in collision in the Tevatron.
- Antiproton Source Improvements: A number of improvements were made to the stochastic cooling systems in the Antiproton Source in order to accommodate higher antiproton flux generated by continuously increasing numbers of protons on the antiproton production target. Improvements included the introduction of transverse stochastic cooling into the Debuncher and upgrades to the bandwidth of the core cooling system. These improvements supported an accumulation rate of $7\times10^{10}$ antiprotons per hour in concert with the above listed improvements.

The above items constituted the improvements associated with Collider Run I. The luminosity goal of Run I was $1\times10^{31}$ cm$^{-2}$sec$^{-1}$, a factor of ten beyond the original Tevatron design goal.

- Main Injector: The Main Injector was designed to significantly improve antiproton performance by replacing the Main Ring with a larger aperture, faster cycling machine [4]. The goal was a factor of three increase in the antiproton accumulation rate (to $2\times10^{11}$ per hour), accompanied by the ability to obtain 80% transmission from the Antiproton Source to the Tevatron from antiproton intensities up to $2\times10^{12}$. An antiproton accumulation rate of $2.5\times10^{11}$ per hour was achieved in Collider Run II, and transmission efficiencies beyond 80% for high antiproton intensities were routine.
- Recycler: The Recycler (see Section 2.3) was added to the Main Injector Project midway through the project (utilizing funds generated from an anticipated cost under run). As conceived the Recycler would provide storage for very large numbers of antiprotons (up to $6\times10^{12}$) and would increase the effective production rate by recapturing unused antiprotons at the end of collider stores [5]. The Recycler was designed with stochastic cooling systems, but R&D on electron cooling was initiated in anticipation of providing improved performance. Antiproton intensities above $5\times10^{12}$ were ultimately achieved although routine operation was eventually optimized around $4\times10^{12}$ antiprotons. Recycling of antiprotons was never implemented for reasons described below.

The Main Injector and Recycler constituted the improvements associated with Collider Run II [6]. The formally established luminosity goal of Run II was $8\times10^{31}$ cm$^{-2}$sec$^{-1}$, a factor of five beyond Run I. However, incorporation of the Recycler into the Main Injector Project was projected to provide up to an additional factor of 2.5.

**1.4 Collider Run I and Run II**

The Run I improvements were all implemented in the early to mid-1990's and supported operations of Collider Run I over the period from August of 1992 through February 1996. Run I consisted of two distinct phases, Run Ia which ended in May of 1993, and Run Ib which was initiated in December of 1993. The 400 MeV linac upgrade was implemented between Run Ia and Run Ib. Run I ultimately delivered a total integrated luminosity of 180 pb$^{-1}$ to both CDF and D0 experiments at $\sqrt{s}$ = 1800 GeV. By the end of the run the typical luminosity at the beginning of a store was about $1.6\times10^{31}$ cm$^{-2}$sec$^{-1}$, a 60% increase over the Run I goal. (A brief colliding beam run at $\sqrt{s}$ = 630 GeV also occurred in Run I.)

The Main Injector and Recycler were completed in the spring of 1999 with the Main Injector initially utilized in the last Tevatron fixed target run. Run II was initiated in March of 2001 and continued through 2011. A number of difficulties were experienced in the initial years of operations. These were ultimately overcome through a lot of experience accumulated in the course of operation and the organization and execution of a "Run II Upgrade Plan". A critical element in the evolution of Run II was the successful introduction of electron cooling [7] into the Recycler in the summer of 2005. Prior to electron cooling luminosities had barely exceeded $1.2\times10^{32}$ cm$^{-2}$sec$^{-1}$, as had been anticipated with the design of the Main Injector. The success of electron cooling supported typical Tevatron luminosities well in excess of $3\times10^{32}$ cm$^{-2}$ sec$^{-1}$, with record stores exceeding $4.3\times10^{32}$ cm$^{-2}$sec$^{-1}$.

While the ultimate performance of Run II exceeded expectations by roughly 70%, the means of achieving this performance differed from the initial plan. In particular, antiproton recycling (the recovery of unspent antiprotons at the end of stores) was never implemented. Difficulties in the removal of protons at 980 GeV in the Tevatron prior to antiproton deceleration proved problematic, and the stunning success of electron cooling and an optimization of store duration time removed the imperative for antiproton recycling. Also, a plan to implement 132 nsec bunch spacing [8], allowing operations with roughly 100 bunches was eventually abandoned. The primary motivation for 132 nsec was to limit the number of interactions per crossing in the detectors to roughly 2-3 as the luminosity increased. However, the utilization of 132 nsec would have required the introduction of a crossing angle in the Tevatron and a corresponding reduction in luminosity of roughly a factor of two. More decisively, the CDF and D0 experiments developed methods for dealing with the higher number of interactions per crossing (up to an average 10 interactions/crossing) without compromising their performance.

## 2. Fermilab Accelerator Complex Operation

The accelerator complex at Fermilab supports a broad physics program including Tevatron Collider Run 2, neutrino experiments using 8 GeV and 120 GeV proton beams, as well as a test beam facility and other fixed target experiments using 120 GeV primary proton beams. This section provides a brief description of each of the accelerators in the chain as they operate at the end of the Collider Run II (2011) and an outline of the Collider shot-setup process (cycle of injection, acceleration, low-beta squeeze, and collisions).

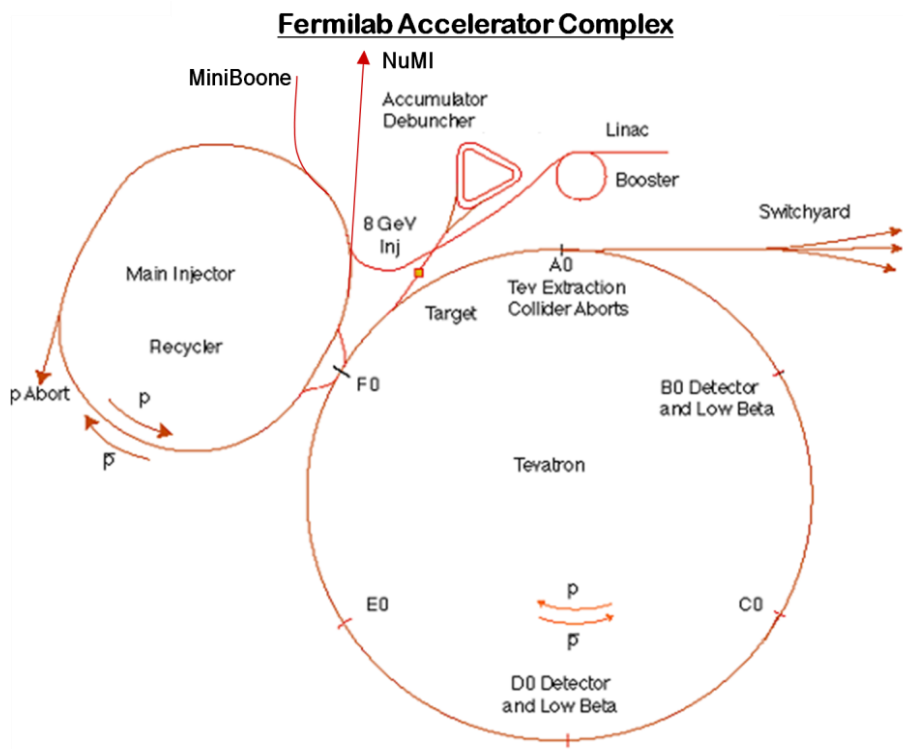

Figure 1: Layout of the Fermilab accelerator complex.

### 2.1 Proton Source

The Proton Source consists of the Pre-Accelerator (Pre-Acc), Linac, and Booster. For operational redundancy, there are two independent 750 kV Pre-Acc systems which provide $H^-$ ions for acceleration through the Linac. Each Pre-Acc is a Cockroft-Walton accelerator having its own magnetron-based $H^-$ source running at a 15 Hz repetition rate, a 750 kV Haefely voltage multiplier to generate the 750 kV accelerating voltage, and chopper to set the beam pulse length going into the Linac. The typical $H^-$ source output current is 50-60 mA.

The Linac accelerates $H^-$ ions from 750 keV to 400 MeV. Originally, the Linac was a 200 MeV machine made entirely of Alvarez-style drift tube tanks [9], but a 1991 upgrade replaced some of the drift tubes with side coupled cavities to allow acceleration up to 400 MeV [10]. Today, the low energy section (up to 116 MeV) is made of drift tube tanks operating with 201 MHz RF fed from tetrode-based 5 MW power amplifier tubes. The high energy section (116 – 400 MeV) consists of 7 side-coupled cavities powered by 805 MHz 12 MW klystrons running giving providing a gradient of ≈7 MV/m. A transition section between the two linac sections provides the optics matching and rebunching into the higher frequency RF system. The nominal beam current out of the Linac is 34 mA.

The Booster is a 474 meter circumference, rapid-cycling synchrotron ramping from 400 MeV to 8 GeV at a 15 Hz repetition rate. (Note that while the magnets ramp at 15 Hz, beam is not present on every cycle.) Multi-turn injection is achieved by passing the incoming $H^-$ ions through 1.5 μm thick (300 μg/cm$^2$) carbon stripping foils as they merge with the circulating proton beam on a common orbit. The 96 10-foot long combined-function Booster gradient magnets are grouped into 24 identical periods in a FOFDOOD lattice [11]. The Booster RF system (harmonic number = 84) consists of 19 cavities (18 operational + 1 spare) that must sweep from 37.9-52.8 MHz as the beam velocity increases during acceleration. The ferrite tuners and power amplifiers are mounted on the cavities in the tunnel. The cavities provide a sum of ≈750 kV for acceleration. The Booster transition energy is 4.2 GeV which occurs at 17 ms in the cycle. Figure 2 illustrates Booster efficiency for various beam intensities during an acceleration cycle; the efficiency is 85-90% for typical beam intensities of 4.5-5.0×10$^{12}$ protons per pulse. Collimators are used localize as much of the lost beam as possible to reduce the radiation dose absorbed by technicians during maintenance periods. Figure 3 shows how Booster throughput has increased remarkably over the Booster operational history. A majority of the proton flux through Booster is delivered to the 8 GeV and 120 GeV neutrino production targets.

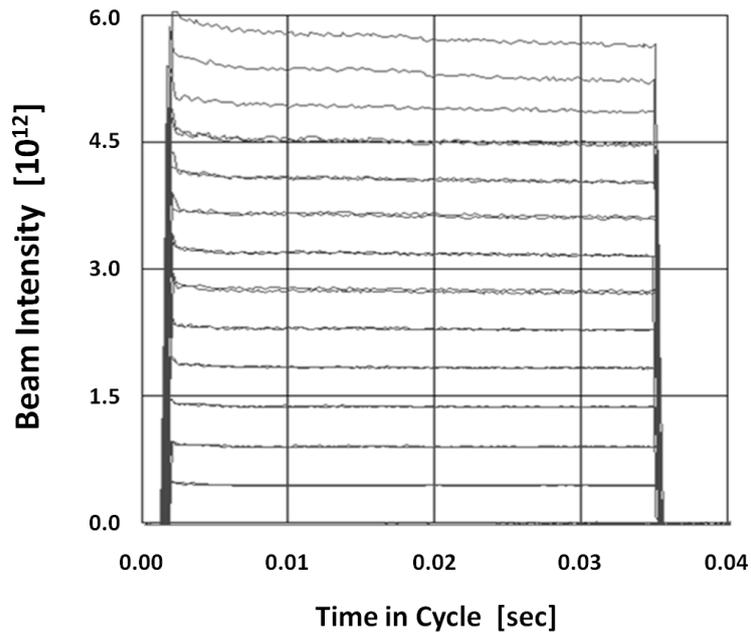

Figure 2: Booster proton intensity for various beam intensities in an acceleration cycle. The injected beam intensity varies from 1 to 13 turns of Linac pulses. The average efficiency is ~90%.

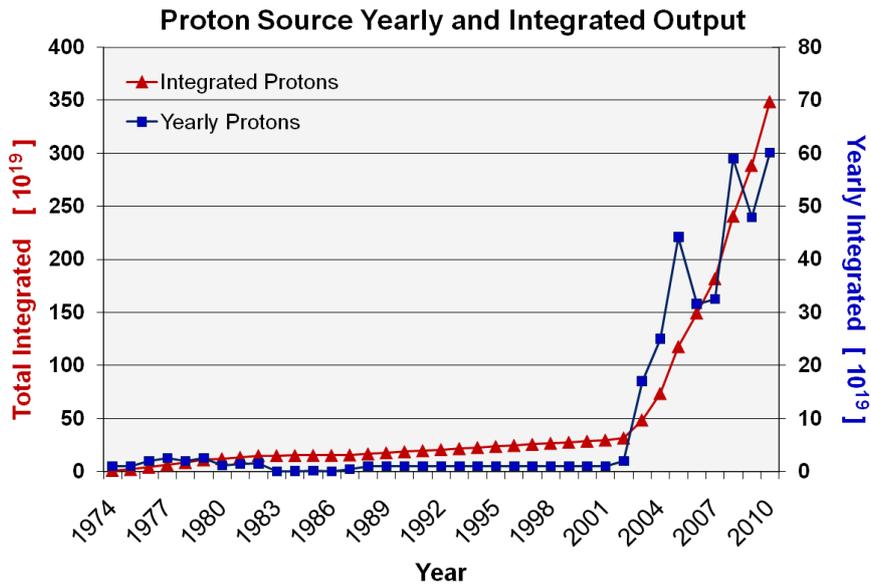

Figure 3: Yearly and integrated proton flux from the Fermilab Booster. The sharp increase in 2003 corresponds to the initiation of the 8 GeV neutrino program.

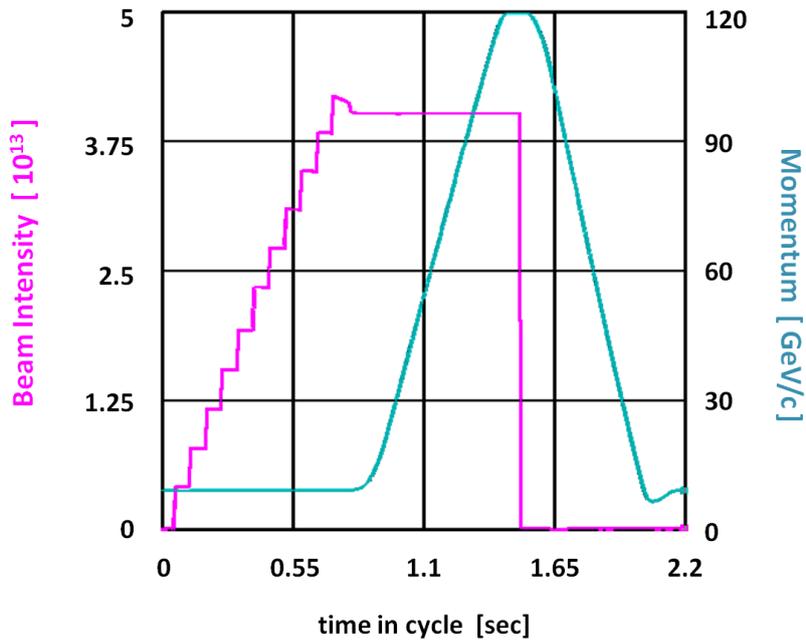

Figure 4: A 120 GeV Main Injector cycle illustrating 11 batch proton injection and acceleration.

**2.2 Main Injector**

The Main Injector (MI) [4] is 3319.4 m in circumference and can accelerate beam from 8 GeV up to 150 GeV. It has a FODO lattice using conventional, separated function dipole and quadrupole magnets. There are also trim dipole and quadrupoles, skew quadrupole, sextupole and octupole magnets in the lattice. Since the Main Injector circumference is seven times the Booster circumference, beam from multiple consecutive Booster cycles, called batches, can be injected around the Main Injector. In addition, even higher beam intensity can be accelerated by injecting more than seven Booster batches through the process of slip-stacking: capturing one set of injected proton batches with one RF system, decelerating them slightly, then capturing another set of proton injections with another independent RF system, and merging them prior to acceleration. There are 18 53 MHz RF cavities are grouped into 2 independently controlled systems to allow slip-stacking and flexibility when an RF station is switched off for maintenance. Beam-loading compensation and active damping systems have been implemented to help maintain beam stability. For beam injections into the Tevatron, coalescing of several 53 MHz bunches of protons and antiprotons into single, high intensity bunches also requires 2.5 MHz system for bunch rotations and a 106 MHz cavity to flatten the potential when recapturing beam into the single 53 MHz bunch. Like the Booster, a set of collimators was installed in the Main Injector to help localize beam losses to reduce widespread activation of components that technicians need to maintain.

The Main Injector supports various operational modes for delivering beam across the complex. For antiproton and neutrino production, 11 proton batches from Booster are injected and slip-stacked prior to acceleration. After reaching 120 GeV, 2 batches are extracted to the antiproton production target while the remaining 9 batches are extracted to the NuMI neutrino production target – see Figure 4. At peak performance, the Main Injector can sustain 400 kW delivery of 120 GeV proton beam power for 2.2 sec cycle times. The Main Injector also provides 120 GeV protons in a 4 sec long slow-spill extracted to the Switchyard as primary beam or for production of secondary and tertiary beams for the Meson Test Beam Facility and other fixed-target experiments. In addition, the Main Injector serves as an effective transport line for 8 GeV antiprotons being transferred from the Accumulator to the Recycler for later use in the Tevatron. Protons from Booster and antiprotons from Recycler are accelerated up to 150 GeV in the Main Injector and coalesced into higher intensity bunches for injection into the Tevatron for a colliding beam store.

**2.3 Recycler**

The Recycler [5] is a permanent magnet 8 GeV/c storage ring whose components are hung from the ceiling above the Main Injector. The Recycler is used as an intermediate storage ring for accumulating significantly larger stashes of antiprotons than can be accommodated in the Antiproton Accumulator. The main Recycler magnets are combined-function strontium ferrite permanent magnets arranged in a FODO lattice. Powered trim magnets are used to make orbit and lattice corrections. An important feature of the Recycler is an electron cooling section to augment its stochastic cooling of the antiprotons. A Pelletron (electrostatic accelerator of the Van-der-Graaf type) provides a 4.3 MeV electron beam (up to 500 mA) that overlaps the 8 GeV antiprotons in a 20 m long section and cools the antiprotons both transversely and longitudinally. After becoming operational in September 2005, electron cooling in the Recycler allowed significant improvements in Tevatron luminosity by providing higher intensity antiprotons with smaller emittances. With electron cooling, the Recycler has been able to store up to $600 \times 10^{10}$ antiprotons. In routine operation, the Recycler accumulates $350\text{-}450 \times 10^{10}$ antiprotons with ~200 hr lifetime for injection to the Tevatron.

**2.4 Antiproton Source**

The Antiproton Source [12] has 3 main parts: the Target Station, the Debuncher, and the Accumulator (AA). Each of these is described briefly below while outlining the steps of an antiproton production cycle. In the Target Station batches of 120 GeV protons ($8 \times 10^{12}$ per batch) transported from the Main Injector strike one of the Inconel (a nickel-iron alloy) layers of the target every 2.2 sec. The beam spot on the target can be controlled by a set of quadrupole magnets. The target is rotated between beam pulses to spread depletion and damage uniformly around the circumference. The shower of secondary particles emanating from the target are focused both horizontally and vertically by a pulsed, high current lithium lens that can provide up 1000 T/m gradient. Downstream of the Li lens is a pulsed dipole magnet that steers negatively-charged particles with 8 GeV/c momentum into the AP2 transport line toward the Debuncher. A collimator between the lens and pulsed magnet was installed to help protect the pulsed magnet from radiation damage as the incoming primary proton beam intensity increased with proton slip-stacking in Main Injector.

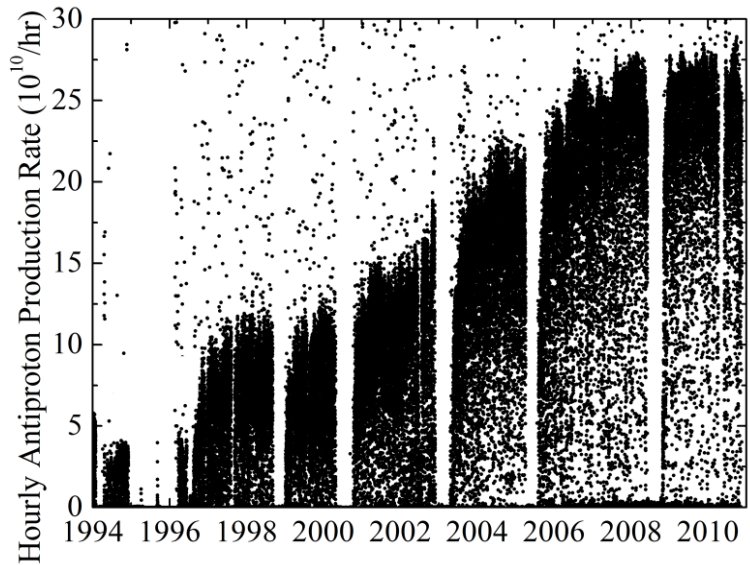

Figure 5: Average antiproton accumulation rate since 1994 and during all of Collider Run II (including production in the Antiproton Source and storage in the Recycler).

The Debuncher and Accumulator are both triangular-shaped rings of conventional magnets sharing the same tunnel. While the Debuncher has a FODO lattice, the Accumulator lattice has particular features needed for cooling and accumulating antiprotons with stochastic cooling systems. The ~$2\times10^8$ antiprotons entering the Debuncher from the AP2 line retain the 53 MHz bunch structure from the primary protons that struck the production target. A 53 MHz RF system (harmonic number = 90) is used for bunch rotation and debunching the antiprotons into a continuous beam with low momentum spread. An independent 2.4 MHz RF system provides a barrier bucket to allow a gap for extraction to the Accumulator. Stochastic cooling systems reduce the transverse emittance from 30 to 3 $\pi$ mm-mrad (rms, normalized) and momentum spread (95% values) from 0.30% to <0.14% prior to injection into the Accumulator.

In the Accumulator, antiprotons are momentum-stacked and cooled by a series of RF manipulations and stochastic cooling. The incoming antiprotons are captured and decelerated 60 MeV by a 53 MHz RF system (harmonic number = 84) to the central orbit where the beam is adiabatically debunched. Before the next pulse of antiprotons arrives (every 2.2 sec), the so-called stacktail momentum stochastic cooling system decelerates the antiprotons another 150 MeV toward the core orbit where another set of independent betatron and momentum stochastic cooling systems provides additional cooling while building a "stack" of antiprotons. Figure 5 shows the average antiproton accumulation rates since 1994; typical values for recent Run II operation are in the range 24-26 $\times10^{10}$/hr.

## 2.5 Tevatron

The Tevatron proton-antiproton collider is a 1-km radius superconducting magnet synchrotron [1, 6] that accelerates beam from 150 to 980 GeV and provides head-on collisions at two interaction points for the CDF and D0 experimental detectors. The Tevatron magnets, arranged in a FODO lattice, are single-bore, warm-iron wound with Nb-Ti superconductor and cooled to 4.5K with liquid helium. A common bus powers the main dipoles and quadrupoles. Other magnetic elements, such as dipole and quad correctors, skew quads, normal and skew sextupoles, and octupoles are located in so-called "spool-packages" adjacent to the lattice quadrupoles. The interaction points are centered within straight sections with dedicated low-beta quadrupole triplets that can squeeze the beams to a ß* of 28 cm. Since both protons and antiprotons circulate in a single beampipe, electrostatic separators are used to kick the beams onto separate helical orbits. Both beams have 36 bunches - 3 trains of 12 bunches with 396 ns spacing (corresponding to 21 buckets of the 53 MHz RF system.) The eight RF cavities in the Tevatron are phased to provide independent control of the protons and antiproton beams. A two-stage collimation system (tungsten primary scatterers and stainless steel secondary absorbers) are used to reduce backgrounds from beam halo in the experiments.

A typical collider fill cycle is shown in Figure 6. First, proton bunches are injected one at a time on the central orbit. Then, electrostatic separators are powered to put the protons onto a helical orbit. Antiproton bunches are injected four bunches at a time into gaps between the three proton bunch trains. After each group of 3 antiproton transfers, the gaps are cleared for the subsequent set of transfers by "cogging" the antiprotons – changing the antiproton RF cavity frequency to let them slip longitudinally relative to the protons. One the beam loading is complete, the beams are accelerated to the top energy (86 sec) and the machine optics is changed to the collision configuration in 25 steps over 125 sec (low-beta squeeze). The last two stages include initiating collisions at the two collision points and removing halo by moving in the collimators. The experiments then commence data acquisition for the duration of the high-energy physics (HEP) store.

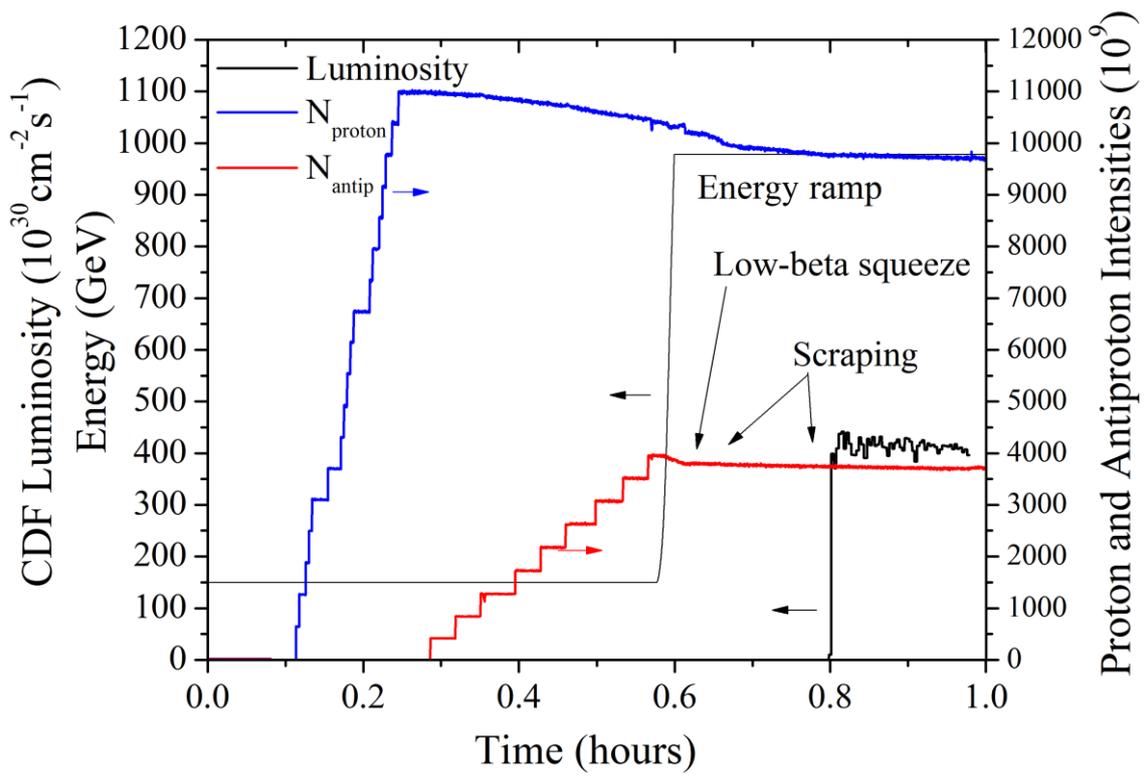

Figure 6: Collider fill cycle for store 8709 (May 2011).

## 3. Collider Performance

Table II summarizes the performance achievements of Run I and Run II. Figure III-1 displays the history of luminosity performance for Runs I and II. Performance in Run II ultimately exceeded the original Tevatron Collider goal by a factor of more than 400.

The Collider performance history – see Figure 7 – shows that the luminosity increases occurred after numerous improvements, some of which were implemented during operation and others introduced during regular shutdown periods. The major improvements of the Collider Run II are listed in the Table III. They took place in all accelerators of the Collider complex and addressed all parameters affecting luminosity – proton and antiproton intensities, emittances, optics functions, bunch length, losses, reliability and availability, etc. – and led to fractional increases varying from few % to some 40% with respect to previously achieved level. There were also other changes and upgrades to hardware and software whose benefit is difficult to quantify, but they made some of the listed improvements possible or easier to implement. One such example is an upgrade to the electronics of the Tevatron Beam Position Monitor (BPM) system that improved the position resolution from ~100 μm to ~10 μm. This upgrade allowed us to see subtle orbit changes during HEP stores that were not visible previously, and a new online orbit correction scheme was developed to improve orbit stability. The BPM electronics upgrade also allowed more accurate optics measurements and, consequently, improved optics corrections implemented in the Tevatron lattice. Another improvement was the installation of additional Tevatron electrostatic separators that increased the separation of the proton and antiproton beams, especially at the parasitic crossing points nearest the main interaction points, and increased the luminosity lifetime by ~15%.

Table II: Design and achieved performance parameters for Collider Runs I and II (typical values at the beginning of a store).

|  | Run Ia | Run Ib | Run II |  |
|---|---|---|---|---|
| Energy (center-of-mass) | 1800 | 1800 | 1960 | GeV |
| Protons/bunch | 1.2 | 2.3 | 2.9 | $\times 10^{11}$ |
| Antiprotons/bunch | 3.1 | 5.5 | 8.1 | $\times 10^{10}$ |
| Bunches/beam | 6 | 6 | 36 |  |
| Total Antiprotons | 19 | 33 | 290 | $\times 10^{10}$ |
| Proton emittance (rms, normalized) | 3.3 | 3.8 | 3.0 | π mm-mrad |
| Antiproton emittance (rms, normalized) | 2 | 2.1 | 1.5 | π mm-mrad |
| β* | 35 | 35 | 28 | cm |
| Luminosity (Typical Peak) | 5.4 | 16 | 340 | $\times 10^{30}$ cm$^{-2}$sec$^{-1}$ |
| Luminosity (Design Goal) | 5 | 10 | 200 | $\times 10^{30}$ cm$^{-2}$sec$^{-1}$ |

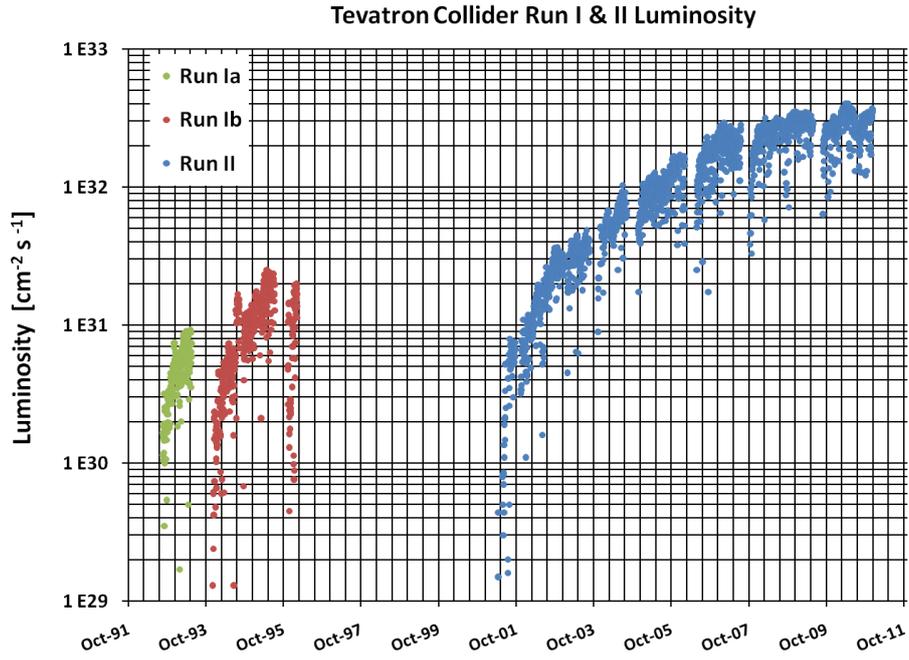

Figure 7: Initial luminosity for all stores in Collider Runs I and II

Table III: Tevatron Collider Run II major luminosity improvements history.

| Improvement | | Luminosity gain |
|---|---|---|
| Optics correction in Accumulator (AA) to Main Injector (MI) beam line | 12/2001 | 25% |
| Tevatron quenches on abort stopped by electron lens | 02/2002 | 0%, reliability |
| Antiproton loss at the step #13 of Tevatron low-beta squeeze fixed | 04/2002 | 40% |
| New Tevatron injection helix implemented | 05/2002 | 15% |
| New AA lattice reduces IBS, emittances | 07/2002 | 40% |
| Beam Line Tuner to reduce emittance dilution at Tevatron injection | 09/2002 | 10% |
| Antiproton multi-bunch coalescing efficiency improved in MI | 10/2002 | 5% |
| Small aperture Lambertson magnets removed from Tevatron C0 sector | 02/2003 | 15% |
| Tevatron sextupoles tuned / SEMs taken out of antiproton beam lines | 06/2003 | 10% |
| New Tevatron helix implemented on ramp to reduce beam losses | 08/2003 | 2% |
| Tevatron magnet reshimming (to center coils inside iron yoke) | 12/2003 | 10% |
| MI dampers operations / HEP store length increased | 02/2004 | 30% |
| Improved efficiency of 2.5 MHz antiproton transfers from AA to MI | 04/2004 | 8% |
| Reduction of Tevatron $\beta^*$ to 35 cm | 05/2004 | 20% |
| Antiproton injections from both Recycler and Accumulator | 07/2004 | 8% |
| Electron cooling system in Recycler operational | 01-07/2005 | ~25% |
| Longitudinal slip-stacking system in Main Injector operational | 03/2005 | ~20% |
| Tevatron octupoles optimized at injection energy of 150 GeV | 04/2005 | ~5% |
| Further reduction of the Tevatron beta-function at IPs $\beta^*$ to 28 cm | 09/2005 | ~10 % |
| Antiproton production optimization | 02/2006 | ~10 % |
| Tevatron helical separation scheme at 150 GeV improved, more protons | 06/2006 | ~10 % |
| Tevatron collision helical separation scheme improved, better lifetime | 07/2006 | ~15 % |
| New Recycler working point results in smaller antiproton emittances | 07/2006 | ~25 % |
| Faster antiproton beam transfers from AA to RR (1 hour →1min) | 12/2006 | ~15% |
| New antiproton target with higher gradient Li lens operational | 01/2007 | ~10% |
| Tevatron sextupole magnet circuits set up for new working point | 2007 | ~10% |
| Compensation of $2^{nd}$ order chromaticity in Tevatron beam optics | 2008 | ~5% |
| Shot-setup time reduced by multi-bunch proton injection | 2008-09 | ~5% |
| Better proton beam quality by scraping in Main Injector | 2008 | ~5% |
| Antiproton beam size dilution at collisions / B0 aperture opened up | 2008 | ~5% |
| Booster proton emittances reduced / tune up of P1 and A1 transfer lines | 04/2010 | ~10% |
| Tevatron collimators employed during low-beta squeeze, more protons | 04/2011 | ~8% |

As the result of some 32 improvements in 2001-2011, the peak luminosity has grown by a factor of about 54 from $L_i \approx 8\times10^{30}$ cm$^{-2}$s$^{-1}$ to $L_f \approx 430\times10^{30}$ cm$^{-2}$s$^{-1}$, or about 13% per step on average. The pace of the Tevatron luminosity progress was one of the fastest among high energy colliders [13].

The Tevatron integrated luminosity has greatly progressed over the years, too – see Figure 7 – and as of June 2011, more than 11 fb$^{-1}$ has been delivered to each detector (CDF and D0).

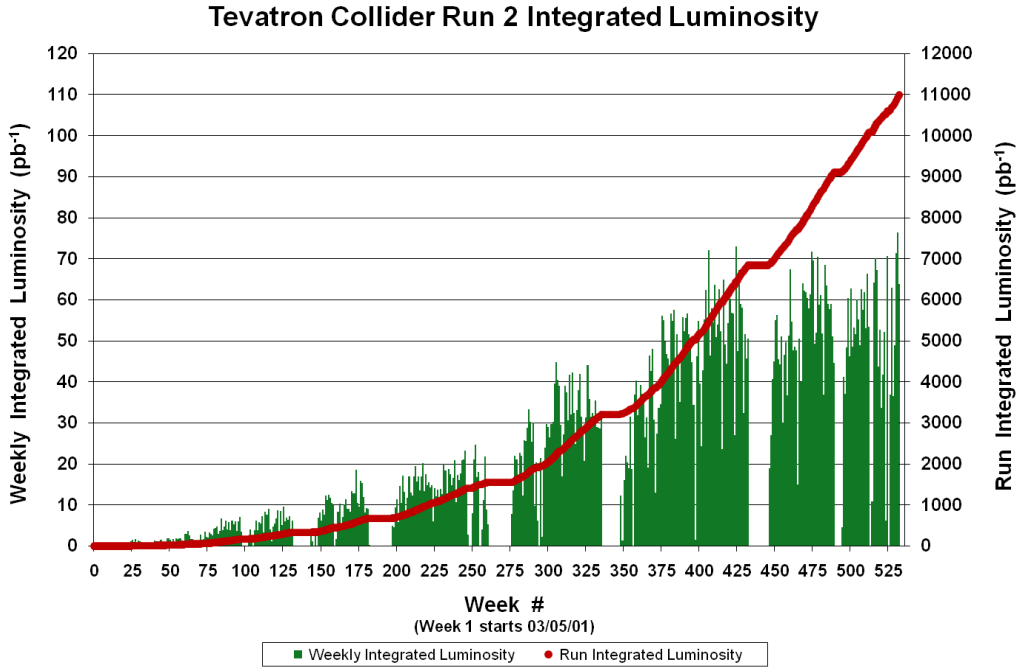

Figure 8: Weekly and total integrated luminosity over Tevatron Run II (2001-2011).

## Acknowledgments

We would like to thank J.Peoples, C.Gattuso, K.Gollwitzer, V.Lebedev and P.Derwent for many useful discussions on the subject of this report.